\documentclass{appolb}
\usepackage{cite}
\usepackage{hyperref}
\usepackage{graphicx}
\usepackage{amsmath}
\begin{document}
\title{Accretion of the relativistic Vlasov gas onto a moving Schwarzschild black hole: Low-temperature limit and numerical aspects
\thanks{Talk presented at the 7th Conference of the Polish Society on Relativity, Łódź, Poland, September 20--23, 2021.}%
}
\author{Patryk Mach and Andrzej Odrzywo{\l}ek
\address{Instytut Fizyki Teoretycznej, Uniwersytet Jagiello\'{n}ski\\ {\L}ojasiewicza 11, 30-348 Kraków, Poland}
\\
}
\maketitle
\begin{abstract}
New developments related to our recent study of the accretion of the Vlasov gas onto a moving Schwarzschild black hole are presented. We discuss the low-temperature limit of the mas accretion rate and a simple Monte Carlo simulation used to check the results obtained in this limit. We also comment on several numerical aspects related with momentum integrals expressing the particle density current and the particle density.
\end{abstract}

\section{Introduction}

Stationary accretion onto a moving black hole is a classic subject dating back to seminal articles by Hoyle, Lyttleton, and Bondi \cite{hoyle_lyttleton, lyttleton_hoyle, bondi_hoyle}. Existing relativistic models are mainly numerical \cite{font, font_ibanez, font_ibanez_papadopoulos, zanotti, blakely, lora, cruz}, but there are also exact analytic results, for instance a solution representing stationary accretion of the ultra-hard perfect fluid onto a moving Kerr black hole \cite{petrich}.

We have recently presented a new exact solution describing a collisionless Vlasov gas accreted by a moving Schwarzschild black hole \cite{maly, duzy}. The gas is assumed to be in thermal equilibrium at infinity, where it obeys the Maxwell-J\"{u}ttner distribution. The techniques used to obtain these solutions were developed by Rioseco and Sarbach in \cite{Olivier1, Olivier2} in the context of a spherically symmetric stationary Bondi-type accretion (see also \cite{sphere} for the case in which boundary conditions are imposed at a finite radius).

This short contribution presents a few new developments and ideas related to the results described in \cite{maly, duzy}. In particular, we discuss the low-temperature limit, in which the flow can also be described by a ballistic approximation derived by Tejeda and Aguayo-Ortiz in \cite{tejeda}.

\section{Accretion of the collisionless Vlasov gas onto a moving Schwarzschild black hole: Solutions}

We work in horizon-penetrating Eddington-Finkelstein spherical coordinates $(t,r,\theta,\varphi)$ and assume a reference frame associated with the moving black hole---at infinity the gas moves uniformly in a given direction with respect to the black hole. The Schwarzschild metric has the form
\begin{equation}
g = - N dt^2 + 2(1 - N) dt dr + (2 - N) dr^2 + r^2 (d \theta^2 + \sin^2 \theta d \varphi^2),
\label{efgeneral}
\end{equation}
where $N = 1 - 2M/r$.

A collisionless Vlasov gas is composed of particles moving along geodesics. It is described in terms of a distribution function $f = f(x^\mu,p_\nu)$, where $x^\mu$ denotes the position, and $p_\mu$ is the four-momentum. The condition that $f$ should be conserved along geodesics yields the Vlasov equation, which we write in the form
\begin{equation}
\label{vlasoveq}
\frac{\partial H}{\partial p_\mu} \frac{\partial f}{\partial x^\mu} - \frac{\partial H}{\partial x^\nu} \frac{\partial f}{\partial p_\nu} = 0,
\end{equation}
where $H(x^\alpha,p_\beta) = \frac{1}{2} g^{\mu \nu}(x^\alpha) p_\mu p_\nu$ is the Hamiltonian corresponding to the geodesic motion. Important observable quantities can be computed as momentum integrals. The particle current density is given by the integral
\begin{equation}
\label{jgeneral}
J_\mu = \int p_\mu f (x,p) \sqrt{- \mathrm{det}[g^{\mu\nu}(x)]} dp_0dp_1dp_2dp_3.
\end{equation}
The particle number density can be defined covariantly as $n = \sqrt{- J_\mu J^\mu}$.

In our analysis we only take into account two types of geodesic orbits. The first class consists of trajectories of particles that originate at infinity and get absorbed by the black hole. They are denoted with the subscript `(abs)'. The second class is formed by trajectories which also originate at infinity, but whose angular momentum is sufficiently high, so that they are scattered by the black hole. These trajectories are denoted with the subscript `(scat)'. Integral quantities such as the particle current density are computed as a sum $J_\mu = J_\mu^\mathrm{(abs)} + J_\mu^\mathrm{(scat)}$.

A stationary solution is specified by asymptotic conditions, which we assume in the form of the Maxwell-J\"{u}ttner distribution corresponding to a simple, i.e., composed of same-mass particles, non-degenerate gas, boosted with a constant velocity $v$ along the axis ($\theta = 0, \theta = \pi$). In the flat Minkowski spacetime it can be written as
\begin{equation}
\label{juttnera}
f(x^\mu,p_\nu) = \delta\left( \sqrt{-p_\mu p^\mu} - m \right) F(x^\mu,p_\nu),
\end{equation}
where
\begin{equation}
\label{boostedf}
F (x^\mu,p_\nu) = \alpha \exp \left\{ \frac{\beta}{m} \gamma \left[ p_t - v \left( \cos \theta p_r - \frac{\sin \theta}{r} p_\theta \right) \right] \right\},
\end{equation}
and the Dirac delta enforces the mass shell condition. Here $\gamma = (1 - v^2)^{-1/2}$ is the Lorentz factor associated with the velocity $v$ and $m$ denotes the rest-mass of a single particle. The constant $\beta$ is related to the temperature $T$ of the gas at rest by $\beta = m/(k_\mathrm{B} T)$, where $k_\mathrm{B}$ denotes the Boltzmann constant.

We treat formulas (\ref{juttnera}) and (\ref{boostedf}) as providing asymptotic conditions for the accretion solution in the Schwarzschild spacetime. The resulting formulas for $J_\mu$ are lengthy and they involve double integrals (see Eqs.\ (29a)--(29d) of \cite{maly} or Eqs.\ (36a)--(36d) of \cite{duzy}). These integrals can be computed numerically leading to solutions of the type shown in Fig.~\ref{Fig:density}.

\section{Monte Carlo and ballistic approximation}

\begin{figure}[t]
\centerline{%
\includegraphics[width=0.8\textwidth]{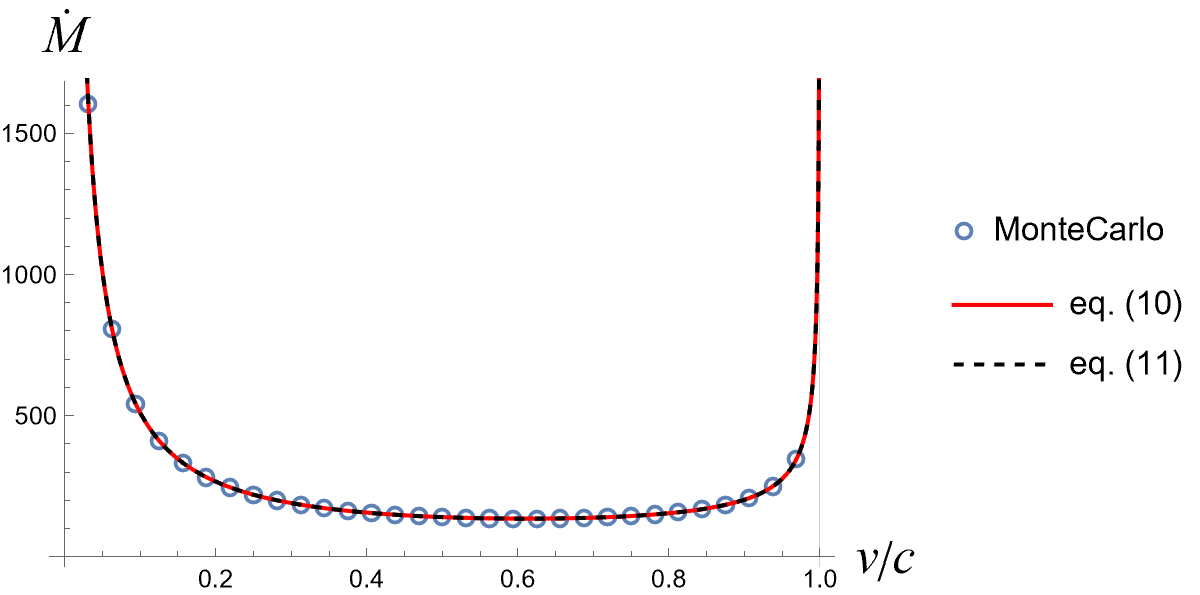}}
\caption{Comparizon of mass accretion rates $\dot M$ obtained in the low-temperature limit. Open circles denote results of the Monte Carlo simulation. Solid red and dotted black lines correspond to Eqs.\ (\ref{formula1}) and (\ref{formula2}), respectively.}
\label{Fig:MonteCarlo}
\end{figure}

A simpler formula can be obtained for the mass accretion rate, defined as
\begin{equation} 
\label{Mdot_general}
\dot M = - m \int_0^{2 \pi} d\varphi \int_0^\pi d \theta\; r^2 \sin \theta \; J^r,
\end{equation}
where the integral is taken over a sphere of constant radius $r$. It reads
\begin{equation}
\label{full_dotM}
\dot M  = \dot M_\mathrm{HL}(1)  \times \frac{\beta}{4 K_2(\beta)} \int_1^\infty d \varepsilon \; e^{-\beta \gamma \varepsilon} \lambda_c^2(\varepsilon)\frac{\sinh \left( \beta \gamma v \sqrt{\varepsilon^2 - 1}  \right)}{\beta \gamma v  \sqrt{\varepsilon^2 - 1}},
\end{equation}
where $K_2$ is the modified Bessel function of the second kind and $\lambda_c$ is defined as
\begin{equation}
\label{lambdac}
\lambda_c(\varepsilon) = \sqrt{\frac{12}{1-\frac{4}{\left(\frac{3 \varepsilon }{\sqrt{9 \varepsilon ^2-8}}+1\right)^2}} }.
\end{equation}
 The dimensional factor $\dot M_\mathrm{HL}(1) = 4 \pi M^2 \rho_\infty$ is the (Newtonian) Hoyle-Lyttleton accretion rate\footnote{In SI units the Hoyle-Lyttleton accretion rate can be expressed as $\dot M_\textrm{HL}(V) = 4 \pi G^2 M^2 \rho_\infty/V^3$, where the uppercase $V \equiv c v$ is the velocity in m/s, $c$ is the speed of light, and $G$ denotes the gravitational constant.}
\begin{equation}
\dot M_\mathrm{HL}(v) = \frac{4 \pi M^2 \rho_\infty}{v^3} 
\end{equation}
corresponding to the speed of light $v = 1$. Here $\rho_\infty = m n_\infty$ denotes the asymptotic rest-mass density of the gas.

In this short contribution we are interested in the low-temperature limit ($\beta \to \infty$). In this limit Eq.\ (\ref{full_dotM}) yields
\begin{equation}
\label{formula1}
\dot{M} = \dot M_\mathrm{HL}(1) \times \frac{\lambda_c(\gamma)^2}{4 \gamma v}
\end{equation}
(see \cite{duzy} for details). For low temperatures thermal motions become negligible, and the particles at infinity move along lines parallel to the axis ($\theta = 0$, $\theta = \pi$). As a consequence, the accretion can be described simply by `shooting' at the black hole with particles moving along such geodesics at infinity. This is the basic idea behind the ballistic approximation discussed in \cite{tejeda}. The resulting formula reads \cite{tejedapriv}
\begin{equation}
\label{formula2}
\dot{M} = \dot M_\mathrm{HL}(1) \times \frac{ \left(8 v^4+20 v^2+\left(8 v^2+1\right)^{3/2}-1\right)}{8 v^3 \sqrt{1-v^2}}.
\end{equation}
Remarkably, expressions \eqref{formula1} and \eqref{formula2} are mathematically equivalent. Note that Eq.\ (\ref{formula2}) differs slightly from Eqs.\ (32) and (33) in \cite{tejeda}. The difference is caused by a different criterion discriminating between scattered and absorbed trajectories used in \cite{tejeda}. 

Both formulas (\ref{formula1}) and (\ref{formula2}) can also be confirmed using a very simple Monte Carlo type simulation. The procedure is as follows. We start with the Schwarzschild metric in a quasi-Cartesian form
\[ g_{\mu \nu} = \left(
\begin{array}{cccc}
 -1 + \frac{2 M}{R} & 0 & 0 & 0 \\
 0 & \frac{R^3-2 M \left(y^2+z^2\right)}{R^2 (R-2 M)} & \frac{2 M x y}{R^2 (R-2 M)} & \frac{2 M x z}{R^2 (R-2 M)} \\
 0 & \frac{2 M x y}{R^2 (R-2 M)} & \frac{R^3-2 M \left(x^2+z^2\right)}{R^2 (R-2 M)} & \frac{2 M y z}{R^2 (R-2 M)} \\
 0 & \frac{2 M x z}{R^2 (R-2 M)} & \frac{2 M y z}{R^2 (R-2 M)} & \frac{R^3-2 M \left(x^2+y^2\right)}{R^2 (R-2 M)} \\
\end{array}
\right), \]
where $R=\sqrt{x^2+y^2+z^2}$ and generate corresponding geodesic equations using the Copernicus Center General Relativity Package for \textit{Mathematica} \cite{ccgrg}. Because the problem is cylindrically symmetric
with respect to, say $z$ axis, we can restrict ourselves  to the $x = 0$ plane. In the next step, we create a bundle of geodesics $x^\mu(s)$ with initial positions $x(0)=0$, $ y(0)=d$, $z(0)  = z_\infty$,
and initial 4-velocities $u^\mu = ( \gamma, 0, 0, - \gamma v )$, where $d$ is random. Geodesic equations are solved numerically, and those geodesics which fall into the black hole are counted as contributing to the accretion rate. In practice, the condition $y(s)^2+z(s)^2 < 8 M^2$ can be used to terminate the integration, as massive particles which fall below the photon sphere $r=3 M$, eventually fall into the black hole. For the Schwarzschild spacetime the procedure can be simplified further, because all particles with initial positions $d$ above some threshold value $d_\ast$ miss the black hole, while those for which $d < d_\ast$ are accreted. The final formula for $\dot{M}$ can be also written as
\[ \dot{M} = \pi d_\ast^2 \gamma v. \]
The threshold value $d_\ast$ can be computed analytically, leading to Tejeda's formula \eqref{formula2} \cite{tejedapriv}. Sample results obtained for $z_\infty = 1000 M$ are depicted in Fig.~\ref{Fig:MonteCarlo} with open circles. The code used for this Monte Carlo type simulation is remarkably simple---the entire code consists of less than 20 \textit{Mathematica} lines.

\section{Numerical aspects}

\begin{figure}[t]
\centerline{%
\includegraphics[width=0.8\textwidth]{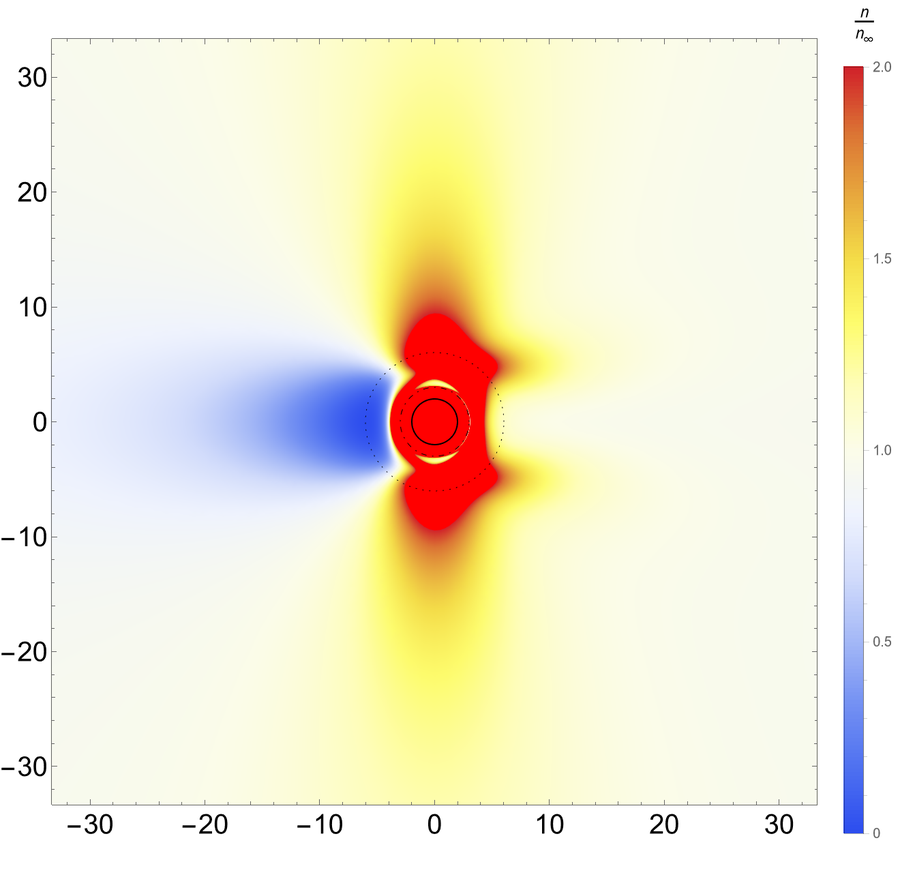}}
\caption{Particle number density map for $\beta=1$ and $v=0.9$. The black hole moves to the right. Note over-density protrusions in front of the black hole, extended disk-like feature perpendicular to the black hole velocity, and a long rarefaction tail behind the black hole. Circles depict the horizon $r=2M$ (solid), the photon sphere $r=3M$ (dot-dashed) and the innermost stable circular orbit $r=6M$ (dotted). Spatial coordinates are expressed in geometrical units ($M = 1$).}
\label{Fig:density}
\end{figure}

In this section we discuss improvements concerning numerical computations of the particle current density $J^\mu$ and the particle density $n$. Originally, several computational hours on a 56-core modern PC were required to create a single image depicting $J^\mu$ or $n$ in \cite{duzy}, similar to Fig.\ \ref{Fig:density}. Calculations were progressively more demanding for increasing values of $v$. The graphs for $v>0.95$ were blurred, indicating poor numerical precision; no results were obtained in a realistic time for $v>0.99$. This forced as to look closer at the implementation details.

Integral expressions for $J^\mu$ derived in \cite{maly, duzy} contain several special functions, which sometimes must be computed by integration as well. An example is provided by the elliptic $X$ function
\begin{equation}
X(\xi,\varepsilon,\lambda) = \lambda \int_\xi^\infty \frac{d \xi^\prime}{{\xi^\prime}^2 \sqrt{\varepsilon^2 - U_\lambda(\xi^\prime)}}.
\label{X}
\end{equation}
While the above integral can be computed analytically, the result is a combination of other elliptic integrals and involves roots of the polynomial appearing in the integrand of \eqref{X}.
A straightforward implementation basing on such expressions works only in \textit{Mathematica}, and it is slow.

To make calculations faster, and to enable an implementation in lower-level programming languages, we created a double exponential integration \cite{dblexp} routine computing $X(\xi,\varepsilon,\lambda)$. It can be written in C or FORTRAN,
and it is effectively $\sim 1000$ x faster than the \textit{Mathematica} internal implementation based on elliptic functions. To our surprise, this improvement had little effect on the overall computational time. Another place in which we hope for a performance improvement is the implementation of modified Bessel functions $I_0$ and $I_1$ appearing in expressions for $J^\mu$. A very fast implementation based on the algorithm described in \cite{I0_I1} is ready but has to be tested.

A 78 x speed-up has been finally achieved in \textit{Mathematica} by a combination of several methods. First, double integration was carefully implemented as nested integrals, avoiding unnecessary error handling and symbolic processing. This required double `encapsulation' of inner integrals, for example as:\\

{\small
\noindent
\pmb{JrInnerAbsF = Function[\{$\xi$, $\theta$, $\varepsilon$\}, NIntegrate[integrand, \{$\lambda$, 0, $\lambda_c$[$\varepsilon$]\}]];}\\
\pmb{JrInnerAbs[$\xi$\_?NumericQ, $\theta$\_?NumericQ, $\varepsilon$\_?NumericQ] :=  JrInnerAbsF[$\xi$, $\theta$, $\varepsilon$];}\\
\pmb{JrAbs = Function[\{$\xi$, $\theta$\}, NIntegrate[JrInnerAbs[$\xi$, $\theta$, $\varepsilon$], \{$\varepsilon$, 1, $\infty$\}]];}\\
}

\noindent Inner functions can be `memoized' \cite{memoization}, but this is pointless in massively-parallel calculations used in practice. Secondly, integration in some regions is unnecessary, as the result is known to be zero. For instance a contribution in $J^\mu$ corresponding to scattered particles vanishes for $r \leq 3M$. Thirdly, the integration in \textit{Mathematica} turns out to be numerically problematic at the horizon and the photon sphere (although the integrands are manifestly regular there). In order to improve performance, we do not evaluate $J^\mu$ at the horizon and the photon sphere. Finally, the requested precision for the outermost integral can be safely reduced from the initial value of $10^{-16}$ to $10^{-4}$.

We expect that the performance improvement achieved so far (by nearly 2 orders of magnitude) should allow for a very detailed analysis of the flow. As an example we present in Fig.\ \ref{Fig:density} a graph of the particle density $n$ obtained for $\beta = 1$ and $v = 0.9$.

We would like to thank Emilio Tejeda for a fruitful correspondence. P.~M.\ was partially supported by the Polish  National Science Centre Grant No.\ 2017/26/A/ST2/00530.


\begin{thebibliography}{99}
\bibitem{hoyle_lyttleton} F.\ Hoyle, R.\ A.\ Lyttleton, The effect of interstellar matter on climatic variation, Proc.\ Cam.\ Phil.\ Soc.\ 35, 405 (1939).

\bibitem{lyttleton_hoyle} R.\ A.\ Lyttleton, F.\ Hoyle, The evolution of the stars, The Observatory, 63, 39 (1940).

\bibitem{bondi_hoyle} H.\ Bondi, F.\ Hoyle, On the mechanism of accretion by stars, Mon.\ Not.\ R.\ Astron.\ Soc.\ 104, 273 (1944).

\bibitem{font} P.\ Papadopoulos, J.\ A.\ Font, Relativistic hydrodynamics around black holes and horizon adapted
coordinate systems, Phys.\ Rev.\ D 58, 024005 (1998).

\bibitem{font_ibanez} J.\ A.\ Font, J.\ M.\ Ib\'{a}\~{n}ez, A numerical study of relativistic Bondi-Hoyle accretion onto a moving black hole: Axisymmetric computations in a Schwarzschild background, Astrophys.\ J.\ 494, 297 (1998).

\bibitem{font_ibanez_papadopoulos} J.\ A.\ Font, J.\ M.\ Ib\'{a}\~{n}ez, P.\ Papadopoulos, Non-axisymmetric relativistic Bondi-Hoyle accretion on to a Kerr black hole, Mon.\ Not.\ R.\ Astron.\ Soc.\ 305, 920 (1999).

\bibitem{zanotti} O.\ Zanotti, C.\ Roedig, L.\ Rezzolla, and L.\ Del Zanna, General relativistic radiation hydrodynamics of accretion flows - I. Bondi-Hoyle accretion, Mon.\ Not.\ R.\ Astron.\ Soc.\ 417, 2899 (2011).

\bibitem{blakely} P.\ M.\ Blakely, N. Nikiforakis, Relativistic Bondi-Hoyle-Lyttleton accretion: A parametric study, Astron.\ Astrophys.\ 583, A90 (2015)

\bibitem{lora} F.\ D.\ Lora-Clavijo, A.\ Cruz-Osorio, and E. Moreno M\'{e}ndez, Relativistic Bondi–Hoyle–Lyttleton accretion onto a rotating black hole: density gradients, The Astrophysical Journal Supplement Series, 219, 30 (2015).

\bibitem{cruz} A.\ Cruz-Osorio, F.\ J.\ S\'{a}nchez-Salcedo, and F.\ D.\ Lora-Clavijo, Relativistic Bondi-Hoyle-Lyttleton accretion in the presence of small rigid bodies around a black hole, Mon.\ Not.\ R.\ Astron.\ Soc.\ 471, 3127 (2017).

\bibitem{petrich} L.\ I.\ Petrich, S.\ L.\ Shapiro, S.\ A.\ Teukolsky, Accretion onto a Moving Black Hole: An Exact Solution, Phys.\ Rev.\ Lett.\ 60, 1781 (1988).

\bibitem{maly} P.\ Mach and A.\ Odrzywo{\l}ek, Accretion of Dark Matter onto a Moving Schwarzschild Black Hole: An Exact Solution, Phys. Rev. Lett. 126, 101104 (2021).

\bibitem{duzy} P.\ Mach and A.\ Odrzywo{\l}ek, Accretion of the relativistic Vlasov gas onto a moving Schwarzschild black hole: Exact solutions, Phys.\ Rev.\ D 103, 024044 (2021).

\bibitem{Olivier1} P.\ Rioseco, O.\ Sarbach, Accretion of a relativistic, collisionless kinetic gas into a Schwarzschild black hole, Class.\ Quantum Grav.\ 34, 095007 (2017).

\bibitem{Olivier2} P.\ Rioseco, O.\ Sarbach, Spherical steady-state accretion of a relativistic collisionless gas into a Schwarzschild black hole, J.\ Phys.\ Conf.\ Ser., 831, 012009 (2017).

\bibitem{sphere} A.\ Gamboa, C.\ Gabarrete, P.\ Dom\'{\i}nguez-Fern\'{a}ndez, D.\ N\'{u}\~{n}ez, and Olivier Sarbach, Accretion of a Vlasov gas onto a black hole from a sphere of finite radius and the role of angular momentum, Phys.\ Rev.\ D 104, 083001 (2021).

\bibitem{tejeda} E.\ Tejeda, A.\ Aguayo-Ortiz, Relativistic wind accretion on to a Schwarzschild black hole, Mon.\ Not.\ R.\ Astron.\ Soc.\ 487, 3607 (2019).

\bibitem{tejedapriv} Emilio Tejeda, \textit{priv. comm} (2021).

\bibitem{ccgrg} A.\ Woszczyna et.\ al., ccgrg - the symbolic tensor analysis package, with tools for general relativity, Wolfram Library Archive (2014), \href{https://library.wolfram.com/infocenter/MathSource/8848/}{https://library.wolfram.com/infocenter/MathSource/8848/}

\bibitem{dblexp} M.\ Mori, Discovery of the Double Exponential Transformation and Its Developments, Publications of the Research Institute for Mathematical Sciences 41, 897 (2005).

\bibitem{I0_I1} P.\ Holoborodko, Advanpix (2015), \href{https://www.advanpix.com/2015/11/11/rational-approximations-for-the-modified-bessel-function-of-the-first-kind-i0-computations-double-precision/}{https://www.advanpix.com/2015/11/11/\\rational-approximations-for-the-modified-bessel-function-of-the-first-kind-i0-computations-double-precision/}

\bibitem{memoization} Wolfram Language \& System Documentation Center (2021),
\href{https://reference.wolfram.com/language/workflow/WriteAFunctionThatRemembersComputedValues.html}{https://reference.wolfram.com/language/workflow/\\WriteAFunctionThatRemembersComputedValues.html}
\end{thebibliography}
\end{document}